\documentstyle[12pt]{article}

\parskip=\medskipamount                 
\def\7#1#2{\mathop{\null#2}\limits^{#1}}        
\def\ast{\displaystyle *}

\def\beee{\begin{equation}}
\def\eeee{\end{equation}}  
\def\dggg{^{\dagger}}     
\begin{document}

\bibliographystyle{unsrt}                                                    

\begin{center}
{{\Large \bf STUDY OF A MODEL OF QUANTUM ELECTRODYNAMICS}
\footnote{email address, owgreen@physics.umd.edu.}}\\
[5mm]
O.W. Greenberg\\
{\it Center for Theoretical Physics\\
Department of Physics \\
University of Maryland\\
College Park, MD~~20742-4111}\\
University of Maryland Preprint PP-00-020[5mm]\\

\end{center}

\vspace{5mm}

\begin{abstract}

This paper studies the model of the 
quantum electrodynamics (QED)
of a single nonrelativistic electron
due to W. Pauli and M. Fierz and studied further by P. Blanchard.  This model 
exhibits infrared divergence in a very simple context.  The infrared divergence
is associated with the inequivalence of the Hilbert spaces associated with the free
Hamiltonian and with the complete Hamiltonian.  Infrared divergences that are visible
in the perturbative description disappear in the space of the clothed electrons.
In this model
when the Hamiltonian is expressed in terms of the 
``physical'' fields
that create the electron together with its cloud of soft photons
the variational principle suggested earlier can be applied.   At finite time the 
Heisenberg field of the
model acts in the space of the perturbative electron together with a finite number 
of perturbative photons, while the
``physical'' field can be chosen to act in the space of the exact 
(``physical'')
electron eigenstates together with a finite number
of physical photons.  
The space of the physical (or clothed) 
electron states can be chosen to be a Fock space.

\end{abstract}

\section{Introduction}

Kurt Haller has pursued fundamental issues in gauge theories. He has especially
emphasized the importance of gauge invariant formulations of quantum 
electrodynamics and of quantum chromodynamics and of the implementation of
Gauss' law in these theories.  Kurt uses operator
methods rather than the more-popular path integral methods; this
provides a good alternative to the usual point of view.  I am 
happy to dedicate this paper to Kurt Haller.

Although there is a long history of the study of charged particles in gauge theories
\cite{c-e,h-l-m,e-m},
there is still concern that the issue is not understood fully.  For example, in a
recent paper \cite{e-m} the authors state `` the battle is not yet won and one
could, in a provacative way, summarize the situation by saying that the question
`what is an electron in QED' is still open.''  

The purpose of this pedagogical paper is to reconsider a very simple model of QED that 
illustrates one facet
of the infrared issues connected with charged particles.  At the end I will discuss
what may carry over to more realistic models of QED as well as to QED itself.  The
model is due to 
W. Pauli and M. Fierz \cite{p-f}. They studied a simple, exactly soluble 
model of a single nonrelativistic electron in QED.
P. Blanchard \cite{bl} studied this
model in the interaction picture and showed that the transition operator is 
well-defined for finite times, but exhibits an infrared divergence for 
$t \rightarrow \pm \infty$.   This model exhibits in an extremely 
simple context
the divergence associated with 
infinite numbers of soft photons associated with a charged particle.  The utility of 
this model is that explicit formulas valid to all orders in $e$
can be found for the
transformation between the perturbative states and fields and the physical ones.

I reconsider this model using the exact eigenstates of the complete Hamiltonian.  
I find that
the Heisenberg field of the model 
at finite times creates states that, because of the infrared divergence, 
are orthogonal to the exact eigenstates.
When the Hamiltonian is expressed
in terms of a field that creates the exact single-particle eigenstates associated
with an electron in this model, the infrared divergence disappears.  I suggest that 
at least part
of the solution to the infrared problem is to replace the original Heisenberg field
by one that creates the electron with the divergent part of its soft photon cloud.

There are three effects associated with the massless quanta in gauge theories:
(1) the infinite number of low-momentum quanta associated with a charged particle,
(2) the collinearity of low-momentum quanta, and (3) the long-range interaction
due to the exchange of massless quanta between charged particles.  Here I study
only the first of these effects.  

F. Bloch and A. 
Nordsieck \cite{b-n} were the first to show how to remove
infrared divergences by summing the cross sections for scattering into final
states that have a charged particle together with any number of soft photons.
D.R. Yennie, S.C. Frautschi and H. Suura \cite{f-s-y} gave a general discussion of 
the 
removal of the divergences in the
cross sections.  P.P. Kulish and L.D. Faddeev \cite{k-f} showed how to 
remove the divergences in the
S-matrix.  I want to show how to remove the divergences in the operator solution
of the Pauli-Fierz model using modified asymptotic fields.  In the charged 
single-particle
model discussed here, the physical field and the modified asymptotic field are the
same.

First I find the exact single-particle eigenstates of the model, which have coherent 
states of soft perturbative photons.  Secondly, I solve the operator equations of 
motion
of the model and find that the Heisenberg field acts in a space that is
orthogonal to the space of the exact eigenstates.
Thirdly, I find fields that
diagonalize the Hamiltonian.  
Finally I discuss which of the properties of this model can be expected to be
relevant to more realistic models of QED as well as to QED itself.

The exact eigenstates in this model 
should have the usual orthonormality properties and thus
the operators that create and annihilate the exact eigenstates should have the
usual free field commutation relations.  This is guaranteed since the transformation 
between these two sets of fields
is formally unitary.  Thus,
following Blanchard \cite{bl} and Contopanagos and
Einhorn \cite{c-e} I assume that the original field operators act in a Hilbert
space associated with coherent states that is unitarily inequivalent to the usual 
Fock space,
while the physical fields that
create the particles together with their soft photon clouds are free fields acting
in the usual Fock space. 
From the point of view of the usual canonical formalism this is a controversial
choice.  The implication of this choice is that in gauge theories and perhaps
in other theories with massless particles or fields the states of relevance 
to observation
are very distant from the states created by smeared polynomials in the fields
in the original Hamiltonian or Lagrangian.
We already know that in field theory the interacting fields create much more than
just the particle with which they are associated.  Further in theories with
infinite field strength renormalization there is an infinite multiplicative 
constant relating the unrenormalized fields, which naively obey canonical
commutation relations, and the renormalized fields whose equal-time commutators
are ill-defined.  The coherent state operator for soft photons that relates the
original fields of the Hamiltonian to the fields that make the single-particle
eigenstates in the Pauli-Fierz model can be viewed as an operator-valued 
analog of the 
field strength renormalization.

Another purpose of this paper is to find the proper extension of the variational
principle proposed earlier \cite{var} to gauge theories.  I will argue that 
in the Pauli-Fierz model, one should first transform to the fields that incorporate
the divergent part of the soft photon cloud associated with the electron in
the exact single-particle eigenstates and then apply the non-gauge
theory form of the variational principle.  

\section{Exact eigenstates in the Pauli-Fierz model}

The Pauli-Fierz model is nonrelativistic quantum electrodynamics in the transverse
gauge with the following approximations: (1)
momentum conservation between the photon and the electron is 
neglected\footnote{This approximation is also made in \cite{k-f}},  
(2) the term quadratic in the vector potential is dropped, and (3) the Coulomb
interaction between electrons is dropped.  So the model is a single electron 
interacting with massless transverse photons.  The Hamiltonian is
\begin{eqnarray}
H &=& \psi\dggg(x)( -\frac{1}{2m}\nabla^2 + V(x)) \psi(x) + \sum^2_{s=1}
\int d^3k k a\dggg_s(k) a_s(k) \nonumber  \\
& &-\frac{e}{m} \sum^2_{s=1} \int \psi\dggg(p) \psi(p) 
\frac{d^3k}{\sqrt{2k}}\tilde{\rho}(k) 
p \cdot e_s(k)
[a_s(k)+a\dggg_s(k)],                                       \label{ham}
\end{eqnarray}
$\tilde{\rho}(k)$ is a smooth function that prevents ultraviolet divergences but
does not affect infrared divergences, in particular $\tilde{\rho}(0)=1$,
and the fields are at time $0$.
For much of the discussion I will suppress
the external potential $V$.

It is straightforward to see that the exact single-electron eigenstates are
\beee
|P\rangle=\sqrt{Z(p)} exp\{\frac{e}{m} \sum^2_{s=1}\int d^3 k 
\frac{\tilde{\rho(k)}}
{\sqrt{2k}k} p \cdot e_s(k) a\dggg_s(k)\} \psi\dggg(p)|0\rangle,  \label{eig}
\eeee
$\psi(p)$ is the Fourier transform of $\psi(x)$, 
with energy
\beee
[\frac{1}{2m}-\frac{2 e^2}{3 m^2} \int_0^{\infty} dk
 \tilde{\rho}^2(k)]p^2 \equiv \frac{p^2}{2m(\infty)}. \label{e}
\eeee
(Fields and states in the physical space will be labelled with capital letters;
vectors such as $p$ will be without bold type or arrows.)
The field strength renormalization factor is 
\beee
\frac{1}{Z(p)}=exp[\frac{e^2 p^2}{m^2} \int_0^{\infty}\frac{dk}{k}
 \tilde{\rho}^2(k)]
\eeee
At first sight the transformation between $|P\rangle$ and 
$\psi\dggg(p)|0\rangle$ does not seem to be even formally unitary.  This first 
impression is misleading and is connected with the fact that annihilation operators
disappear when acting on the vacuum.  In fact the transformation between these
states is
\beee
U=exp\{\frac{e}{m}\int d^3p \psi\dggg(p) \psi(p) d^3k
\frac{\tilde{\rho}(k)}{\sqrt{2k}k} \sum_s p \cdot e_s(k) 
(a\dggg(k)-a(k))\}.                                      \label{unitary}
\eeee
Using the Baker-Haussdorff theorem, we find
\begin{eqnarray}
U & = & exp\{-\frac{e^2p^2}{2m^2}\int_0^{\infty} \frac{dk~\tilde{\rho}^2(k)}{k}\}  
exp\{\frac{e}{m} \int \frac{d^3k~\tilde{\rho}(k)}{\sqrt{2k}k}p 
\cdot e^s(k)a_s\dggg(k)\}  \nonumber \\
  &   &  exp\{-\frac{e}{m}\int \frac{d^3k ~
\tilde{\rho}^2(k)}{\sqrt{2k}k}p \cdot e^s(k)a_s(k)\}
\end{eqnarray}
Clearly when acting on the vacuum the factor with the annihilation operators
becomes one and the remaining factors are not unitary.
If one works in the algebra of operators rather
than in the Hilbert space, one does not lose the annihilation parts of operators.

\section{Calculation of the Heisenberg operator}

The Heisenberg equations of motion are 
\beee
-i\partial_t \psi \dggg(p,t)=\frac{p^2}{2m} \psi \dggg(p,t)
-\frac{e}{m}\sum_{s=1}^2\int\frac{d^3k}{\sqrt{2k}}{\tilde \rho}(k)
\psi \dggg(p,t) p \cdot e_s(k) (a_s(k,t)+a_s\dggg(k,t))   \label{heis}
\eeee
\beee
-i\partial_t a_s\dggg(k,t)=ka_s\dggg(k,t)-\frac{e}{m}\int \frac{d^3p}{\sqrt{2k}}
\tilde {\rho}(k)\psi \dggg(p,t)\psi(p,t) p \cdot e_s(k)  \label{heis-a}
\eeee
Since the Hamiltonian commutes with the product $\psi \dggg(p,t)\psi(p,t)$,
this product is conserved and the time can be dropped in the product when
it appears in Eq.({\ref{heis-a}}).  Then the equation for $a\dggg$ can be solved
exactly,
\beee
a_s\dggg(k,t)=e^{ikt} a_s\dggg(k)-i\frac{e}{m}\int \frac{d^3p}{\sqrt{2k}}p \cdot 
e_s(k)
\psi\dggg(p) \psi(p) \tilde {\rho}(k) t.
\eeee
The term in $a\dggg$ linear is $t$ is connected with the neglect of momentum
conservation in the model.  This linear term cancels in the sum $a\dggg + a$
that appears in the equation of motion for $\psi \dggg$.  The solution for
$\Psi \dggg$ is
\beee
\Psi \dggg(p,t)=C exp\{i \frac{p^2}{2m}t- \frac{e}{m} \sum_{s=1}^2 \int 
\frac{d^3k}{\sqrt{2k}k}\tilde{\rho}(k) p \cdot e_s(k)
[e^{ikt}a_s\dggg(k)-e^{-ikt}a_s(k)]\}.
\eeee
Requiring $\Psi\dggg(p,0)=\psi(p)$ gives
\beee
\Psi \dggg(p,t)=\psi\dggg(p) exp\{i \frac{p^2}{2m}t- \frac{e}{m} \sum_{s=1}^2 \int 
\frac{d^3k}{\sqrt{2k}k}[f_s(p,k,t) a_s\dggg(k)-f_s^{\ast}(p,k,t) a_s(k)]\},
\eeee
\beee
f_s(p,k,t)=\frac{\tilde{\rho}(p)}{\sqrt{2k}k} p \cdot e_s(k)(e^{ikt}-1).
\eeee
The expectation of the number operator $N=\sum_s \int d^3k a\dggg(k) a(k)$ for 
photons in the one electron state is
\beee
<N(t)>=\frac{e^2}{m^2} \sum_s \int d^3k |f_s(k,p,t)|^2   \nonumber
\eeee
\beee
 = \frac{e^2}{m^2} \sum_s \frac{d^3k}{2k^3} \tilde{\rho}^2(k) sin^2\frac{kt}{2}.
\eeee
This is the same expression that Blanchard calls $N^2(t)$.  As Blanchard shows,
for all finite $t$, $<N(t)>$ is finite, but for 
$t \rightarrow \pm \infty$ $<N(t)>$ diverges as $log~ t$.  
Thus for all finite $t$
there is a finite expection value for the 
number of soft photons in the cloud associated with the Heisenberg field of
the electron,
but for $t \rightarrow \pm \infty$ the number diverges.  A divergent value of $<N>$
is the signature for a space of vectors in an inequivalent representation of the
commutation relations that is orthogonal to the original Fock space.  
The fact that
the large-$t$ limit leads out of the usual Fock space has been 
identified as the
cause of infrared divergences by several authors, 
starting with K.O.
Friedrichs \cite{fr} and Blanchard \cite{bl}.  

The expectation value of the number
operator also diverges in the exact eigenstates, 
Eq.(\ref{eig}).  Thus the Heisenberg field acting on the vacuum creates states
that are orthogonal to all the exact electron eigenstates in the model.
This suggests that in this model one should introduce a field 
that makes
the physical single particle state with its attached soft photons. 
In the single
charged particle sector this should lead to diagonalization of the Hamiltonian.

\section{Diagonalization of the Hamiltonian}

We can invert the transformation $U$ given in Eq.(\ref{unitary}) to express
the Hamiltonian in terms of ``physical''
operators that create the charged particles 
together with their soft photon clouds.  We introduce the physical
photon operators (and their adjoints), $A(k)=U a(k) U \dggg$,
\beee
A(k)=a(k)-\frac{e}{m}\int d^3p \psi \dggg (p) \psi(p) f_s(p,k)    \label{a}
\eeee
as well as the transformed electron operators (and their adjoints),
$\Psi(p)=U \psi(p) U \dggg$,
\beee
\Psi(p)=exp\{ \frac{e}{m} \int d^3k [f_s(p,k) 
a\dggg_s(k)-f_s^{\ast}(p,k)a_s(k)]\}\psi(p).        \label{phys-ops}
\eeee
In terms of the physical electron and photon field operators the Hamiltonian
is
\newpage
\begin{eqnarray}
H= \Psi\dggg(x)( -\frac{1}{2m(\infty)}\nabla^2 + V(x)) 
\Psi(x) + \sum^2_{s=1}
\int d^3k k A\dggg_s(k) A_s(k)  +   \nonumber  \\
+\sum_{s=1}^2 \int d^3p d^3p^{\prime} f_s^{\ast}(p,k) f_s(p^{\prime},k) 
\Psi\dggg(p^{\prime}) \Psi\dggg(p) \Psi(p^{\prime})
 \Psi(p).
\end{eqnarray}
The bilinear terms in $\Psi$ correspond to scattering of the electron 
together with its soft photon cloud in the external potential $V$.  No 
infrared divergences or other infrared effects are visible.  The model is
taken to be a one electron model, so we ignore the quartic terms in 
$\Psi(p)$.

This result is similar to very old work on clothed operators in simple models
of field theory \cite{g-s},
except in full QED we would propose to incorporate only the divergent part of the soft
photons in the electron Heisenberg field, 
rather than eliminating all the trilinear terms in the original Hamiltonian.

\section{Condition for the operator variational principle}

In this model, the physical fields diagonalize the single-particle Hamiltonian.
The asymptotic fields will be those that occur in scattering from the fixed potential
$V(x)$.
Thus the condition to be imposed on the Hamiltonian in
the variational principle \cite{var} based on choosing
elements from the algebra of asymptotic fields should be to make the Hamiltonian
as close to the free field Hamiltonian as possible.  This condition is the analog 
of the condition chosen in
theories without massless particles or fields.

\section{What is relevant to QED?}

Some issues suggested by this simple model may carry over to QED. The first is that
one should express the observables of the theory in terms of a charged field that,
acting on the vacuum, creates a state that is not 
orthogonal to the asymptotic states.  The charged field should contain the soft
photons in the neighborhood of zero energy that are responsible for the infrared
divergence in the number of perturbative photons attached to the charged particle.
This can be done by a formally unitary transformation.  The same unitary 
transformation should be applied to the photon field.  
Secondly, as suggested by Contopanogos and
Einhorn \cite{c-e}, one can take the charged states with their clouds of soft
coherent state photons to be in Fock space and the perturbative states to be in
an inequivalent representation of the commutation relations.

In the present model, if one introduces
the clothed electron operators in the theory the infrared divergences disappear
and in the absence of an external potential the Hamiltonian has free field form
because the physical photons decouple from the physical electrons.
This of course will not happen in QED, nor will there be a discrete mass associated
with the single charged particle.  Since the work of B. Schroer \cite{sc}
on ``infraparticles''
we have known that particles that interact with massless quanta cannot have discrete
mass.

Since in the present model the single-particle Hamiltonian is diagonalized by the
physical field, the variational principle
is trivial in this model.  
When the massless quanta have been absorbed the variational
principle proposed earlier for the case where there are no massless particles or 
fields can be applied.  This also will not apply to QED and shows that significant
modifications must be made in this variational principle before it can be useful
in studying gauge theories. 

\section{Qualitative remarks}

Divergent field strength renormalization occurs both due to ultraviolet effects
and to infrared effects.  In the case of ultraviolet divergences, one introduces
the renormalized fields, for example $\psi_{ren}(p)=Z^{-1/2} \psi(p)$, where
the multiplicative factor is fixed by requiring the coefficient of the 
$\psi^{as}(p)$
term in the Haag expansion to be one; see references in
\cite{var}.  This, together with other renormalization effects makes the
Haag expansion finite.  By
contrast in the case of infrared divergences (suppressing non-infrared effects),
one should not introduce a field strength renormalization; rather one should
exponentiate
the soft photons of near zero energy 
into a new operator-valued multiplicative factor multiplying
the charged field as we have done in Eq.(\ref{phys-ops}).  
In the Pauli-Fierz model this
explicitly removes the infrared divergences and produces a charged field that
acting on the vacuum creates a state that is not orthogonal to all the asymptotic
states.  One can hope that in QED at least part of the infrared divergence can be 
removed using the physical charged field and that this 
field acting on the vacuum will produce asymptotic states.

One could argue \footnote{Xiang-Dong Ji, private communication}
that given a Hamiltonian that has no trilinear term one could introduce
trilinear terms in the transformed Hamiltonian and could then arrange to have
any form of infrared divergence.  My point of view is that, although highly
simplified, this model does come from QED and does capture the type of soft
photon infrared
divergence that occurs in QED.

\section{Outlook for future work}

The idea of introducing physical charged fields that contain the infrared divergent
part of the soft photons should be tried in more realistic models, such as the
asymptotic Hamiltonian of Kulish and Faddeev \cite{k-f}, as well as in other models
that have the same infrared divergences as QED.

{\bf Acknowledgements}

I am happy to thank Ted Jacobson, Xiang-Dong Ji and especially 
Ching-Hung Woo for helpful discussions.

\end{document}